\documentclass[a4paper]{jpconf}
\usepackage[small,bf]{caption}
\usepackage{graphicx}
\begin{document}
\title{Measurements of the correlation between electrons from heavy-flavour hadron decays and light hadrons with ALICE at the LHC}
\author{E. Pereira De Oliveira Filho on behalf of the ALICE collaboration}

\address{Universidade de S\~{a}o Paulo, Departamento de F\'{i}sica Nuclear\\
Travessa R da rua do Mat\~{a}o 187\\
05508900, S\~{a}o Paulo, Brasil
}

\ead{epereira@cern.ch}

\begin{abstract}
In relativistic heavy-ion physics two-particle correlations provide a very useful tool to investigate the Quark-Gluon Plasma (QGP). This observable is sensitive to several of the properties of the QGP such as resonances, interaction of partons with the medium and collective effects (e. g. elliptic flow). In the present work, the correlation function between electrons from heavy-flavour hadron decays and light hadrons was measured in pp and Pb-Pb collisions (central and semi-central). Furthermore, in pp collisions the relative beauty contribution to the total cross section of electrons from heavy-flavour decays was estimated by comparing the measured correlation with Monte-Carlo templates.
\end{abstract}

\section{Introduction}
In relativistic heavy-ion collisions, two-particle correlation is a powerful tool to investigate the properties of the Quark-Gluon Plasma (QGP). This observable is sensitive to the near and away side jet structure, resonances and collective effects \cite{ref1, ref2, ref3}. Which effect dominates the correlation is strongly determined by the kinematic region of the investigated particles, as well as the event class as discussed in \cite{ref3}.\\
Heavy quarks (charm and beauty) are of particular importance in this context. They are often used as probes of the QGP, since they are produced mainly during the initial stages of the collisions through hard parton-parton scattering processes, and subsequently they experience the entire evolution of the system and they might participate in the collective expansion of the medium. One of the common approaches in the study of heavy quarks is the measurement of electrons from the decay of hadrons containing such particles. This is the technique exploited in the present work.\\
In order to address the QGP properties, a given observable is often studied as a function of the event class (centrality) of the collision and in various collision systems. The correlation function of electrons from heavy-flavour decays and light hadrons was measured with ALICE in pp and central and semi-central Pb-Pb collisions. Furthermore, for the case of pp collisions, the relative beauty contribution to the total cross section of electrons from heavy-flavour decays was estimated by comparing the measured correlation function with Monte-Carlo templates.
\section{Electron Identification and Background Reconstruction}
\begin{center}
\begin{figure}[h]
\begin{center}
\includegraphics[scale=0.10]{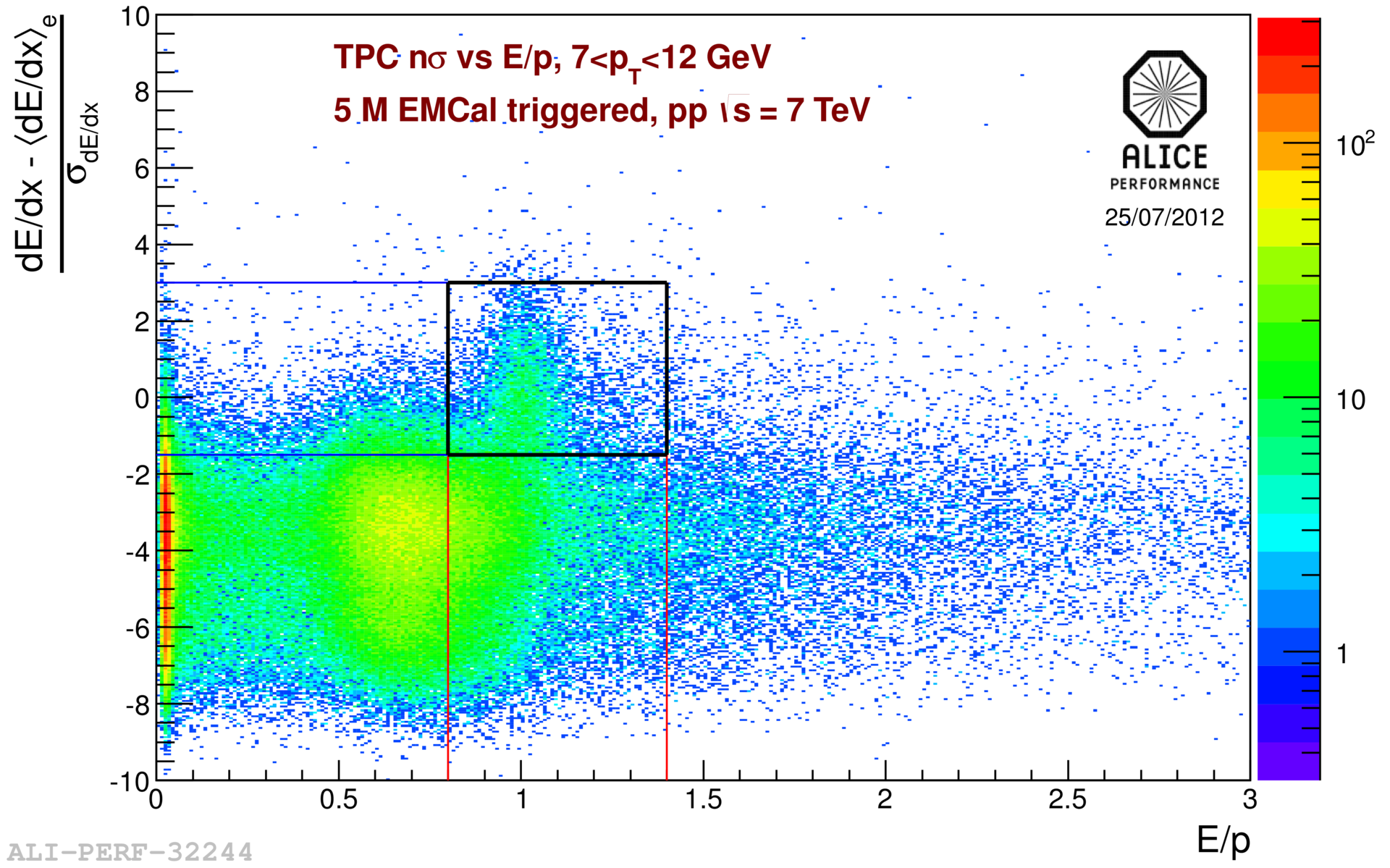} 	
\includegraphics[scale=0.07]{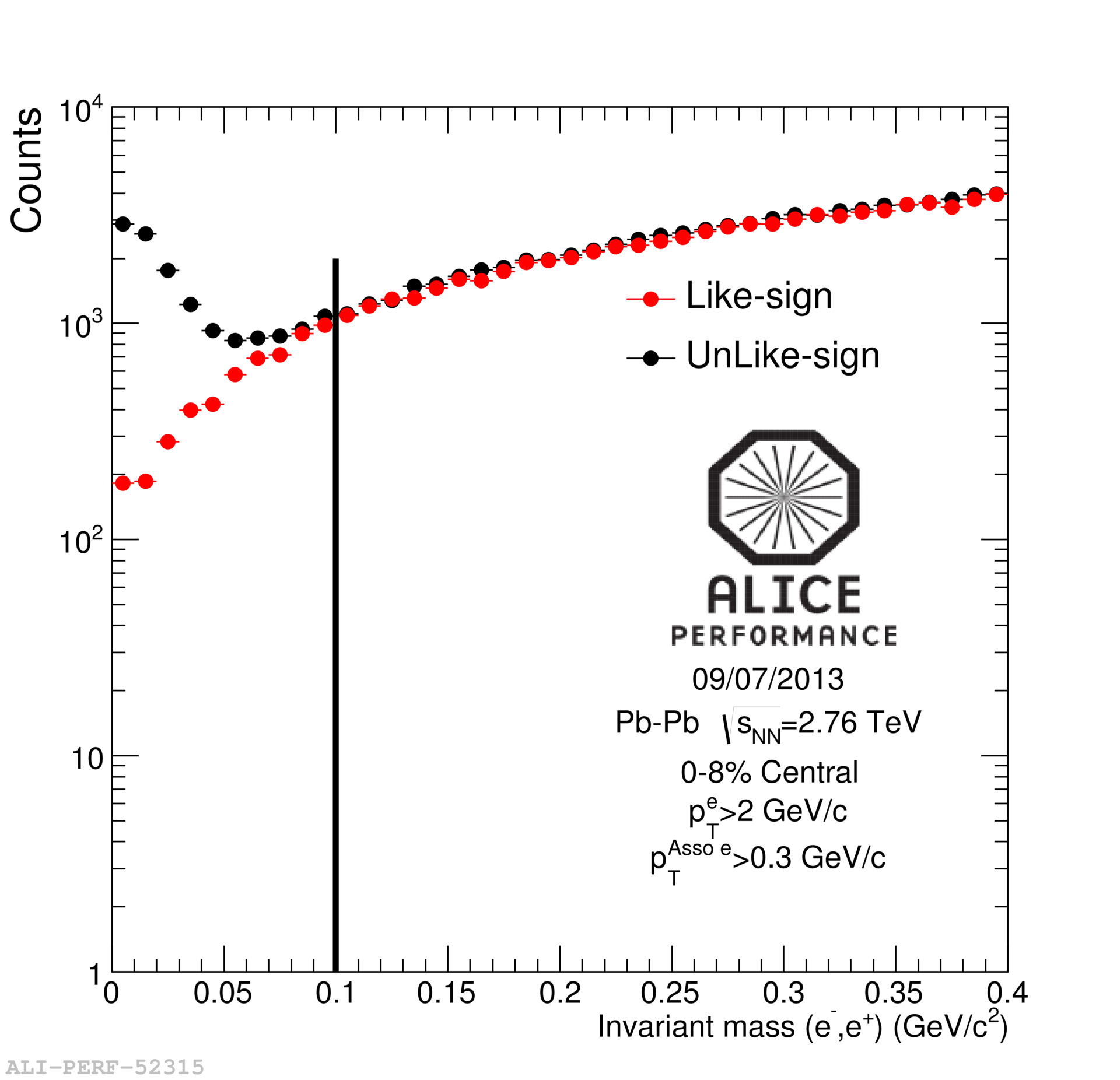} 
\end{center}
\caption{Electron identification using TPC and EMCal information (left). Invariant Mass of $e^{\pm}e^{\mp}$ (black points) and $e^{\pm}e^{\pm}$ (red points) pairs, used to reconstruct electrons from gamma conversions in the detector material and from light meson Dalitz decays (right).}
\label{fig:pid}
\end{figure}
\end{center}

At the LHC, ALICE is dedicated to the study of heavy-ion collisions \cite{ref5}. ALICE has excellent capabilities for heavy-flavour measurements thanks to several subsystems which allow vertexing and track reconstruction with high resolution and particle identification (PID) in a wide momentum range \cite{ref5}. The Time Projection Chamber (TPC) and the ElectroMagnetic Calorimeter (EMCal) are the main PID detectors used in the present analysis. Fig.~\ref{fig:pid} (left) illustrates how electron identification can be performed using TPC and EMCAL (combined) information. The electron band can be seen around $E/p = 1$ and $\rm{TPC} \: N\sigma = 0$, where $E$ is the energy measured in the EMCal and $p$ the track momentum. The $\rm{TPC} \: N\sigma$ is defined as the measured specific energy loss ($dE/dx$)	 relative to the expected value, for a given mass assumption (the electron mass in this case).

After the identification, electrons not coming from heavy-flavour hadron decays should be removed from the sample. This background is mainly composed by electrons from photon conversions in the detector material and light meson (e. g. $\pi^{0}$ and $\eta$ mesons) Dalitz decays. This contribution can be statistically estimated through the invariant mass of $e^{\pm}e^{\mp}$ pairs where the combinatorial background is determined using like-sign pairs ($e^{\pm}e^{\pm}$). Fig.~\ref{fig:pid} (right) shows the background reconstruction in Pb-Pb collisions.

\section{Azimuthal correlation analysis and results}
\begin{center}
\begin{figure}[h]
\begin{center}
\includegraphics[scale=0.134]{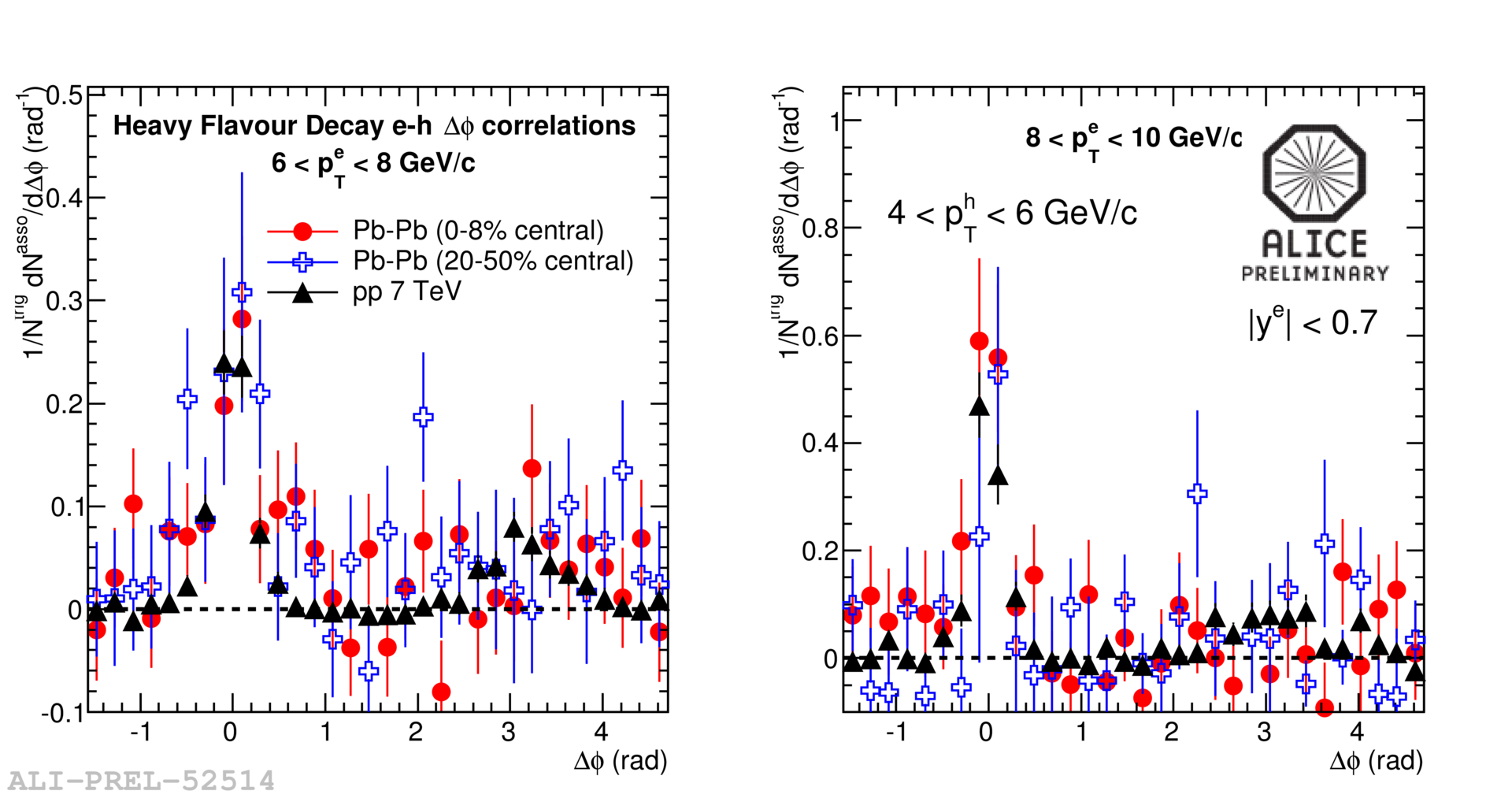} 
\end{center}
\caption{Correlation function between electrons from heavy-flavour decays and light hadrons in pp ($\sqrt{s} = 7 \: \rm{TeV}$) and Pb-Pb ($\sqrt{s_{\rm{NN}}} = 2.76 \: \rm{TeV}$) collisions.\\}
\label{fig:res1}
\end{figure}
\end{center}

\begin{center}
\begin{figure}[h]
\begin{center}
\includegraphics[scale=0.085]{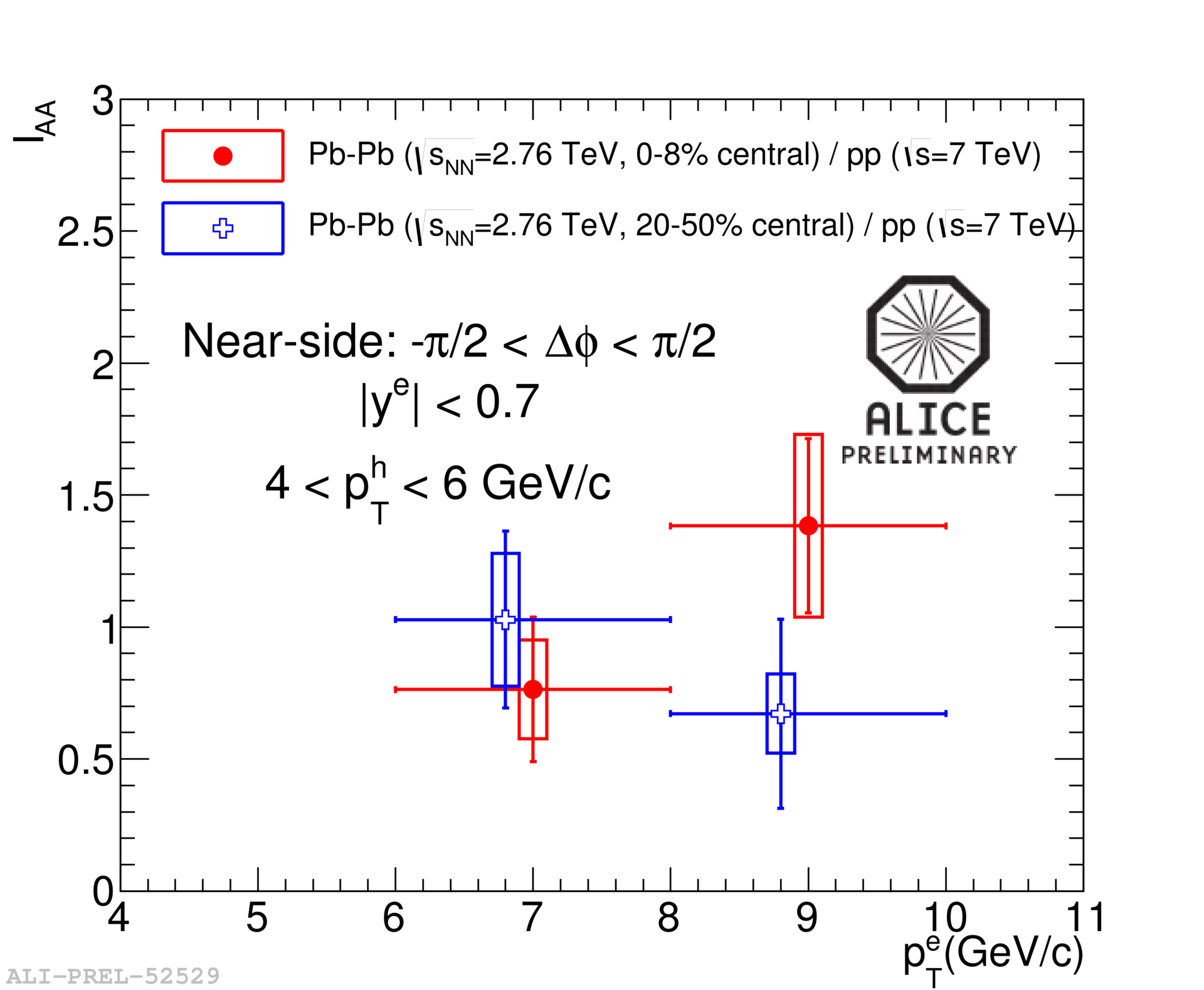} 
\includegraphics[scale=0.085]{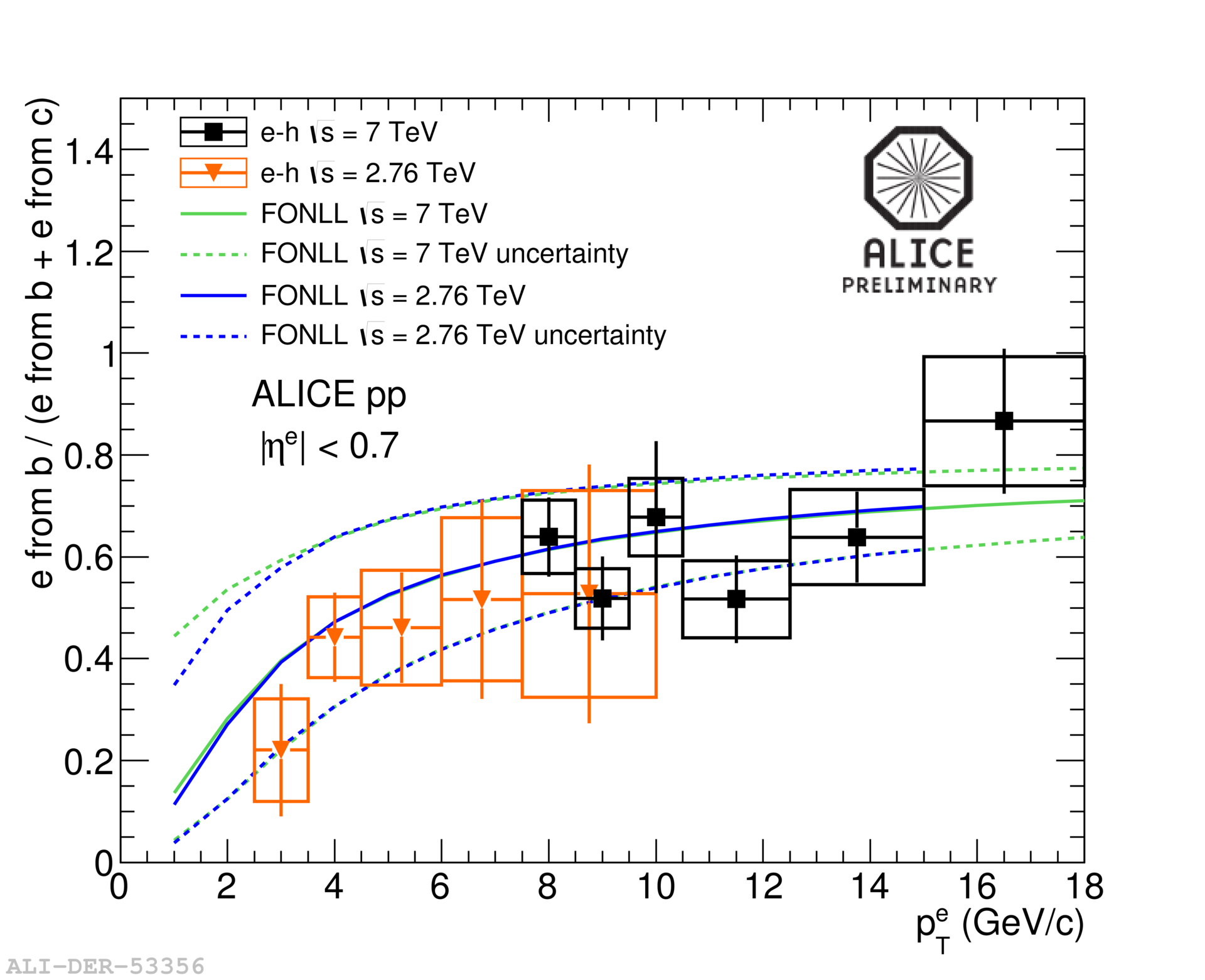} 
\end{center}
\caption{Yield modification in Pb-Pb relative to pp collisions (left). Relative beauty contribution to the total cross section of electrons from heavy-flavour decays, as a function of $p_{\rm{T}}$, in pp collisions (right).}
\label{fig:res2}
\end{figure}
\end{center}

The azimuthal ($\Delta\phi$) correlation function between a trigger and an associated particle is defined in Eq.~\ref{eq:e1} \cite{ref1}. For the case of electrons from heavy-flavour decays, the background is removed as discussed in Section 2. The remaining hadron contamination in the electron sample is removed by scaling down the di-hadron correlation distribution and subtracting it from the one of inclusive electrons. The final correlation function is obtained after correcting for the tracking efficiency and for the Mixed Event Correlation, which accounts for acceptance effects \cite{ref1}.
\begin{equation}
C_{t-a}(\Delta\phi) = \left. \left(\frac{dN_{t-a}}{N_{t-a}d(\Delta\phi)}\right)^{Same-Event} \middle/ \left(\frac{dN_{t-a}}{N_{t-a}d(\Delta\phi)}\right)^{Mixed-Event} \right.
\label{eq:e1}
\end{equation}
Fig.~\ref{fig:res1} shows the correlation function of electrons from heavy-flavour hadrons decay and light hadrons for the kinematic range defined by $6 < p_{\rm{T}}^{e} < 8$ GeV/$c$ and $8 < p_{\rm{T}}^{e} < 10$  GeV/$c$ with $4 < p_{\rm{T}}^{h} < 6$  GeV/$c$, for central and semi-central Pb-Pb collisions at $\sqrt{s_{\rm{NN}}} = 2.76 \: \rm{TeV}$, as well as for pp collisions at $\sqrt{s} = 7 \: \rm{TeV}$. Fig.~\ref{fig:res2} (left) shows the modification $I_{\rm{AA}}$ of the near-side yield in heavy-ion collisions relative to pp collisions defined as:
\begin{equation}
I_{\rm{AA}} (p_{\rm{T}}^{e},p_{\rm{T}}^{h}) = \left. \left(\int\limits_{-\frac{\pi}{2}}^{\frac{\pi}{2}} C(\Delta\phi;p_{\rm{T}}^{e},p_{\rm{T}}^{h}) d(\Delta\phi)\right)^{Pb-Pb} \middle/ \left(\int\limits_{-\frac{\pi}{2}}^{\frac{\pi}{2}} C(\Delta\phi;p_{\rm{T}}^{e},p_{\rm{T}}^{h}) d(\Delta\phi)\right)^{pp}  \right.
\label{eq:iaa}
\end{equation}

The relative beauty contribution to the total cross section of electrons from heavy-flavour decays is shown in Fig. 3 (right), for pp collisions at $\sqrt{s} = 2.76 \: \rm{TeV}$ and $7 \: \rm{TeV}$, obtained by fitting the near-side correlation to Monte-Carlo templates obtained with PYTHIA \cite{ref6}.


\section{Conclusions}
The measured $I_{\rm{AA}}$ is consistent with unity within statistical and systematic uncertainties. Indeed, the small statistics available does not allow stronger conclusions, as well as it does not exclude a difference in the correlation between heavy-ion and pp collisions.\\
The relative beauty contribution to the cross section of electrons from heavy-flavour decays is consistent with FONLL pQCD calculations \cite{ref7} within the uncertainties for both collision energies of the present analysis.
\section*{Acknowledgments}
{\small The author thanks all his colleagues from the ALICE Collaboration, University of S\~{a}o Paulo and Utrecht University, for every help, advice and very useful discussions. He also thanks the "Conselho Nacional de Desenvolvimento Cient\'{i}fico e Tecnol\'{o}gico - CNPq", for the financial support.}

\end{document}